\documentclass[aps,prd,nofootinbib,floatfix,superscriptaddress,10pt]{revtex4-1}
\usepackage{epsfig}
\usepackage{graphicx}
\usepackage{multirow}
\usepackage{latexsym,amsbsy,amsmath,mathtools}
\usepackage{bm}
\usepackage{dcolumn}
\usepackage{booktabs}
\def\vec#1{\boldsymbol{#1}}
\def\llcol#1#2{\tilde{\lambda}_{#1}.\tilde{\lambda}_{#2}}
\def\ka{\left.\mid 11 \right\rangle}
\def\kb{\left.\mid 88 \right\rangle}

\newcommand{\be}{\begin{equation}}
\newcommand{\ee}{\end{equation}}
\newcommand{\bea}{\begin{eqnarray}}
\newcommand{\eea}{\end{eqnarray}}
\newcommand{\beas}{\begin{eqnarray*}}
\newcommand{\eeas}{\end{eqnarray*}}

\begin{document} 
\boldmath \title{Spectroscopy, lifetime and decay modes of the 
$T^-_{bb}$ tetraquark}\unboldmath

\author{E.~Hern\'andez}
\email{gajatee@usal.es}
\affiliation{Departamento de F\'\i sica Fundamental e IUFFyM, Universidad de Salamanca, 37008
Salamanca, Spain}
\author{J.~Vijande}
\email{javier.vijande@uv.es}
\affiliation{Unidad Mixta de Investigaci\'on en Radiof{\'\i}sica e Instrumentaci\'on Nuclear 
en Medicina (IRIMED), Instituto de Investigaci{\'o}n Sanitaria La Fe (IIS-La Fe)-Universitat de Valencia (UV) 
and IFIC (UV-CSIC), 46100 Valencia, Spain}
\author{A.~Valcarce}
\email{valcarce@usal.es}
\affiliation{Departamento de F\'\i sica Fundamental e IUFFyM, Universidad de Salamanca, 37008
Salamanca, Spain}
\author{Jean-Marc~Richard} 
\email{j-m.richard@ipnl.in2p3.fr} 
\affiliation{Universit\'e de Lyon, Institut de Physique Nucl\'eaire de Lyon,
IN2P3-CNRS--UCBL,\\ 
4 rue Enrico Fermi, 69622  Villeurbanne, France} 
\date{\today} 
\begin{abstract}
We present the first full-fledged study of the flavor-exotic isoscalar 
$T_{bb}^-\equiv b b \bar u \bar d$ tetraquark with spin and parity $J^P=1^+$.
We report accurate solutions of the four-body problem in a quark model, characterizing the structure 
of the state as a function  of the ratio $M_Q/m_q$ of the heavy to light quark masses.  
For such a standard constituent model,  $T_{bb}^-$ lies
approximately 150\,MeV below the strong decay threshold $B^-\bar {B^*}^{0}$ and 105\,MeV below 
the electromagnetic decay threshold $B^- \bar B^0 \gamma$. 
We evaluate  the lifetime of $T_{bb}^-$, identifying the promising decay modes where the 
tetraquark might be looked for in future experiments. Its total decay width is
$\Gamma \approx 87 \times 10^{-15}$\,GeV and therefore
its lifetime $\tau \approx$ 7.6 ps. The promising final states are ${B^*}^{-}\, {D^*}^{+} \, \ell^- \, \bar \nu_\ell$
and $\bar {B^*}^{0} \, {D^*}^{0} \, \ell^- \, \bar \nu_\ell $ among the semileptonic decays, and 
${B^*}^{-} \, {D^*}^{+} \, {D_s^*}^-$, $\bar {B^*}^{0} \, {D^*}^{0} \, {D_s^*}^- $, and 
${B^*}^{-} \, {D^*}^{+} \, \rho^-$ among the nonleptonic ones. 
The semileptonic decay to the isoscalar $J^P=0^+$ tetraquark $T_{bc}^0$ is also relevant but
it is not found to be dominant.
There is a broad consensus about the existence of this tetraquark, and its detection 
will validate our understanding of the low-energy realizations of 
Quantum Chromodynamics (QCD) in the multiquark sector. 
\end{abstract}
\maketitle 
\section{Introduction}
Hadronic physics has been much stimulated  during the last two decades by the experimental 
discovery of several  new resonances in the hidden-charm sector, resonances that are hardly accommodated in the 
traditional quark-antiquark or three-quark picture~\cite{Che16}. These are the so-called $XYZ$ mesons 
and LHCb pentaquarks, which belong to the class of ''exotic hadrons'', although they are not flavor exotics. 
After several years of studies, no definite conclusion has  
been drawn as to whether such non-flavor exotic states correspond to multiquark 
structures or to hadron-hadron molecules. A similar situation was encountered 
in the light scalar meson sector, where a multiquark picture
was first introduced~\cite{Jaf76} as an attempt to explain the inverted mass 
spectrum (inverted in comparison to the simple quark-antiquark structure favored by 
the naive quark model) exhibited by the low-lying scalar mesons, some of which were later
on suggested to be meson-meson molecules~\cite{Wei90}.

For years, the sector of flavor-exotic hadrons has been 
somewhat forgotten, as being less easily accessible than the hidden-flavor
sector. However, already some decades ago, investigations on flavor-exotic 
multiquarks concluded that $QQ\bar q \bar q$ four-quark configurations 
become more and more deeply bound when the mass ratio $M_Q/m_q$ increases~\cite{Ade82}.
There is nowadays a broad theoretical consensus about the existence of such unconventional 
tetraquark configurations for which all strong decays are energetically 
forbidden. The most promising candidate is an isoscalar
tetraquark with double beauty and  $J^P=1^+$ quantum numbers, 
which is stable against strong and electromagnetic decays. The same conclusion 
about the stability of this state has been reached in a wide variety of theoretical 
approaches~\cite{Ade82,Eic17,Kar17,Fra17,Bic16,Jun19,Vij09,Ric18,Luo17,Duc13,Cza18}. 
A novel lattice QCD calculation~\cite{Fra17} employing a non-relativistic formulation 
to simulate the bottom quark finds unambiguous signals for a strong-interaction-stable 
$(I)J^P=(0)1^+$ tetraquark, 189(10)\,MeV below the corresponding
two-meson threshold, $\bar B\bar B^*$. The lattice QCD calculation of Ref.~\cite{Jun19} 
comes to the identical
  conclusion obtaining a binding energy of 143(34)\,MeV. With such binding, the tetraquark is stable 
also with respect to electromagnetic decays. In Ref.~\cite{Kar17}, the mass of the 
doubly-charm baryon $\Xi_{cc}^{++}$, discovered by the LHCb Collaboration~\cite{Aai17}, is used 
to calibrate the binding energy of a $QQ$ diquark. Assuming that 
the $bb$ diquark binding energy in a $T^-_{bb}$ tetraquark is the same as 
that of the $cc$ diquark in the $\Xi_{cc}^{++}$ baryon,
the mass of the $(0)1^+$  doubly-bottom tetraquark is estimated to be 215\,MeV below 
the strong decay threshold $\bar B\bar B^*$. 
In Ref.~\cite{Eic17}, the heavy-quark-symmetry mass relations linking heavy-light 
and doubly-heavy-light mesons and baryons are combined with leading-order corrections for
finite heavy-quark mass, corresponding to hyperfine spin-dependent terms and kinetic energy 
shift that depends only on the light degrees of freedom. This leads to  predict   that the
$T^-_{bb}$ state is stable against strong decays. More specifically, using as input 
the masses of the doubly-bottom baryons (not yet experimentally measured) obtained by
the model calculations of Ref.~\cite{Kar17}, Ref.~\cite{Eic17} finds an axial-vector 
tetraquark bound by 121\,MeV. 
In Ref.~\cite{Bic16}, the Schr\"odinger equation is solved with a potential extracted 
from a lattice QCD calculation for static heavy quarks, in a regime where the pion mass 
is  $m_\pi\sim 340$\,MeV, and again, evidence is found  for a stable isoscalar 
doubly-bottom axial-vector  tetraquark. When extrapolated to physical pion masses 
it has a binding energy of $90^{+43}_{-36}$\,MeV. 
The robustness of these predictions is reinforced by detailed 
few-body calculations using phenomenological constituent models based 
on quark-quark Cornell-type interactions~\cite{Vij09,Ric18}, which predict 
that the isoscalar axial-vector doubly-bottom tetraquark is strong- and 
electromagnetic-interaction stable with a binding energy ranging between 144--214\,MeV 
for different realistic quark-quark potentials. Recent studies using a simple 
color-magnetic model have come to similar conclusions~\cite{Luo17}. The QCD sum rule analysis 
of Ref.~\cite{Duc13} also points to the possibility of a stable doubly-bottom isoscalar axial-vector 
tetraquark. Finally, the recent phenomenological analysis of Ref.~\cite{Cza18} also presents
evidence in favor of the existence of a stable $T^-_{bb}$ state. 

The compelling theoretical evidence for the existence of a $T^-_{bb}$ tetraquark has led to preliminary
studies of its lifetime and weak decay modes. In this context, Ref.~\cite{Xin18}  
investigated the amplitudes and decay widths of doubly-heavy
tetraquarks under the flavor SU(3) symmetry, deriving
ratios between decay widths of different channels.
Ref.~\cite{Aga19} evaluated the semileptonic decay of the $T^-_{bb}$
   tetraquark to a scalar $bc\bar u\bar d$ tetraquark in the framework of QCD
   sum rules. In spite of the numerous model calculations existing in the literature (see Refs.~\cite{Ric18,Luo17} for a
recent compendium) no comprehensive  calculation of the 
spectroscopy, decay modes, and lifetime of this state has been obtained so far.

The purpose of  this work is to present the first detailed
study of the flavor-exotic  $T_{bb}^-$ tetraquark with $J^P=1^+$ and isospin $I=0$, 
reporting accurate solutions of the four-body problem, characterizing the structure 
of the state, evaluating its lifetime and identifying
the promising decay modes where the tetraquark might be looked for.
A striking result deals with the lifetime, which is found significantly longer 
than for single-$b$ hadrons. With two $b$ quarks, one expects either cooperating 
or conflicting interferences. Also, if one compares a typical meson decay mode 
$B\to D\, x$ and its tetraquark analog $T\to B\, D\, x$, there is a change in the overlap 
of the final $D$ meson and the $c\bar q$ system provided by a spectator $\bar q$ 
and the $c$ quark coming from one of the $b$ quarks: the color factor and the spatial 
distribution are modified. There is also an obvious effect of the phase-space, which 
is known to be crucial in weak decays, for instance for $\beta$-unstable nuclei. While 
the $Dx$ invariant mass is 5.3\,GeV in $B\to Dx$ decay, it is about 4.6\,GeV in 
$T\to  BDx$, depending which sector of the Dalitz plot is reached.  Altogether, 
it looks difficult to attempt a guesstimate of the lifetime before actually performing the calculation. 

This paper is organized as follows. In Sec.~\ref{secII}, we present the masses and 
wave functions obtained from an accurate four-body calculation that makes use of a 
quark model. The calculation of the dominant decay modes and of the lifetime is 
given is Sec.~\ref{secIII}. Our conclusions are summarized in Sec.~\ref{secIV}.

\section{Tetraquark mass and wave function}
\label{secII}
We have studied the spectroscopy of doubly-heavy tetraquarks by two different 
numerical methods: a hyperspherical harmonic formalism and a generalized Gaussian 
variational (GGV) approach, both driving to the same results~\cite{Vij09}. For its later 
application to  the detailed study of the four-quark structure and weak decays,
the GGV is more suited. Let us briefly discuss the main characteristics of 
the method. We shall denote the heavy quark coordinates by $\vec r_1$ and $\vec r_2$,
and those of the light antiquarks by $\vec r_3$ and $\vec r_4$. The tetraquark wave function 
is taken to be a sum over all allowed channels with well-defined symmetry properties~\cite{Via09,Vin09}: 
\be
\psi(\vec x,\vec y,\vec z) = \sum^{6}_{\kappa=1} \chi^{csf}_{\kappa} \, 
R_{\kappa}(\vec x,\vec y,\vec z),
\label{trial-wave-function}
\ee
where $\vec x = \vec r_1 - \vec r_2$, $\vec y = \vec r_3 - \vec r_4$ and $\vec z =(\vec r_1 + \vec r_2
- \vec r_3 - \vec r_4)/2$ are the Jacobi coordinates. $\chi^{csf}_\kappa$ are orthonormalized 
color-spin-flavor vectors and $R_{\kappa}(\vec x,\vec y,\vec z)$ is the radial part 
of the wave function of the $\kappa^\text{th}$ channel. 
In order to get the appropriate symmetry properties
in configuration space, $R_{\kappa}(\vec x,\vec y,\vec z)$ 
is expressed as the sum of four components, 
\bea
R_{\kappa}(\vec x,\vec y,\vec z)&=& \sum^4_{n=1} w_\kappa^n \, R^n_\kappa(\vec x,\vec y,\vec z) ,
\label{Rk}
\eea
where $w_\kappa^n = \pm 1$. Finally, each $ R^n_\kappa(\vec x,\vec y,\vec z)$ is expanded in terms 
of $N$ generalized Gaussians
\bea
R^n_\kappa(\vec x,\vec y,\vec z) = \sum^{N}_{i=1} \alpha^i_{\kappa} \,
\exp\left[ - a^{i}_\kappa\,\vec x^{2} - b^{i}_\kappa \,\vec y^{2} - c^{i}_\kappa \,\vec z^{2}
- d^{i}_\kappa \, s_1(n) \,\vec x\cdot\vec y - e^{i}_\kappa \, s_2(n) \,\vec x\cdot\vec z
- f^{i}_\kappa \, s_3(n) \,\vec y\cdot\vec z\right]\, ,
\label{Rkr}
\eea
where $s_i(n)$ are equal to $\pm 1$ to guarantee the symmetry properties of the radial
function and $\alpha^{i}_\kappa,a^{i}_\kappa,\cdots, f^{i}_\kappa$ are the variational parameters. 
The latter are determined by minimizing the intrinsic energy of the tetraquark. We follow closely
the developments of Refs.~\cite{Via09,Vin09}, where further technical details can be found about the wave function and the minimization procedure. 

A four-quark state is stable under the strong interaction if its mass, 
$M_{T}$ (from now on, $T$ often abbreviates $T_{QQ}$), lies below 
all allowed two-meson decay thresholds. Thus, one can define the difference 
between the mass of the tetraquark and that of the lowest two-meson threshold, 
namely:
\be
\Delta E = M_{T} - (M_{1} + M_{2})\,,
\label{Delta-E}
\ee
where $M_1$ and $M_2$ are the masses of the mesons constituting the threshold.
When $\Delta E < 0$, all fall-apart decays are forbidden and, therefore, the state is stable under
strong interactions. When $\Delta E \geq 0$ one has 
to examine whether it is a resonance or an artifact of the discretization of 
the continuum by the variational method, and this requires dedicated  techniques 
such as real~\cite{Hiy18} or complex scaling~\cite{Oka19}, which are beyond the scope 
of this note. We therefore concentrate on $\Delta E < 0$. Another quantity of interest is the root-mean-square 
(r.m.s.) radius of the tetraquark, $X_T$, given by~\cite{Vij09}:
\be
X_T = \left[\frac{\sum^4_{i=1} m_i \langle (\vec r_i - \vec R)^2 \rangle}{\sum^4_{i=1} m_i} 
\right]^{1/2} \, ,
\label{RMS}
\ee
where $\vec R$ is the center-of-mass coordinate, and $m_i$ are the quark masses $M_Q$ or $m_q$.

Determining whether stability is reached in this model, i.e., $\Delta E<0$, 
requires a simultaneous and consistent calculation of the meson masses $M_1$ 
and $M_2$ entering the threshold, and of the tetraquark configurations.
For this purpose, we have adopted the so-called AL1 model by Semay and 
Silvestre-Brac~\cite{Sem94}, already used in a number of exploratory studies of multiquark 
systems, for instance in our recent investigation of the hidden-charm pentaquark sector
$\bar c c qqq$~\cite{Ric17} or doubly-heavy 
baryons and tetraquarks~\cite{Ric18}. 
It includes a standard Coulomb-plus-linear central potential, supplemented 
by a smeared version of the chromomagnetic interaction,
\begin{eqnarray}
V(r)  & = &  -\frac{3}{16}\, \llcol{i}{j}
\left[\lambda\, r - \frac{\kappa}{r}-\Lambda + \frac{V_{SS}(r)}{m_i \, m_j}  \, \vec \sigma_i \cdot \vec \sigma_j\right] \, , \nonumber\\
V_{SS}(r)  & = & \frac{2 \, \pi\, \kappa^\prime}{3\,\pi^{3/2}\, r_0^3} \,\exp\left(- \frac{r^2}{r_0^2}\right) ~,\quad
 r_0 =  A \left(\frac{2 m_i m_j}{m_i+m_j}\right)^{-B}\!,
\label{eq:mod1}
 \end{eqnarray}
where
$\lambda=$ 0.1653\,GeV$^2$, $\Lambda=$ 0.8321\,GeV, $\kappa=$ 0.5069, $\kappa^\prime=$ 1.8609,
$A=$ 1.6553\,GeV$^{B-1}$, $B=$\,0.2204, $m_u=m_d=$ 0.315\,GeV, $m_s=$ 0.577\,GeV, $m_c=$ 1.836\,GeV and $m_b=$ 5.227\,GeV. 
Here, $\llcol{i}{j}$ is a color factor, suitably modified for the quark-antiquark pairs.
We disregard the small three-body term of this model used in~\cite{Sem94} to fine-tune the baryon 
masses vs. the meson masses. Note that the smearing parameter of the spin-spin term is adapted to 
the masses involved in the quark-quark or quark-antiquark pairs. It is worth to emphasize that the 
parameters of the AL1 potential are constrained in a simultaneous fit of 36 well-established meson 
states and 53 baryons, with a remarkable agreement with data, as seen in Table 2 of Ref.~\cite{Sem94}.
\begin{table}[t]
\caption{Relevant meson masses (in MeV) and r.m.s. radii (in fm) predicted by the AL1 model for 
the strong, electromagnetic and weak decay thresholds of the $J^P=1^+$ $T^-_{bb}$ tetraquark.}
  \begin{tabular}{cp{.5cm}cccp{0.5cm}ccc}
\hline\hline
                         & & Meson     &   M  & r.m.s.&& Meson    &  M   & r.m.s\\ \hline\addlinespace[0.5mm]
\multirow{3}{*}{$J=0$}   & &$\bar B$   & 5293 & 0.145 && $K$      & 491  & 0.283 \\
                         & &$D$        & 1862 & 0.216 && $\pi$    & 138  & 0.298 \\
		                     & &$D_s$      & 1962 & 0.213 && $\eta_c$ & 3005 & 0.181 \\  
\hline \addlinespace[0.5mm] 
\multirow{3}{*}{$J=1$}   & &$\bar B^*$ & 5350 & 0.153 && $K^*$    & 903  & 0.389 \\
                         & &$D^*$      & 2016 & 0.248 && $\rho$   & 770  & 0.460 \\
			                   & &$D^*_s$    & 2102 & 0.243 && $J/\Psi$ & 3101 & 0.199 \\
												\hline\hline
  \end{tabular}
	\label{Tab1}
\end{table}
\begin{table}[b]
\caption{Properties of the $T_{QQ}$ tetraquark as a function of the mass of the
heavy quark $M_Q$ for the AL1 model. Energies and masses are in MeV and distances in fm. 
For $\Delta E>0$, the probabilities and average distances are just an indication that 
the variational calculation will likely not converge toward a bound state.}
\begin{ruledtabular}
  \begin{tabular}{crrrccccccrrrr}
$M_Q$  &
$M_1+M_2$\hspace*{-5pt} &
\multicolumn{1}{c}{$M_{T_{QQ}}$} &
\multicolumn{1}{c}{$\Delta E$}    &
$P[| \bar 3 3\rangle]$ & $P[| 6 \bar 6\rangle]$ &
$P[\ka]$ & $P[\kb]$ & $P_{MM^*}$ &
$P_{M^*M^*}$ &
$\langle x^2\rangle^{1/2}$\hspace*{-5pt} &
$\langle y^2\rangle^{1/2}$\hspace*{-5pt} &
$\langle z^2\rangle^{1/2}$\hspace*{-5pt} &
$X_{T_{QQ}}$\ \ \\ \hline
5227	& 10644	& 10493	   & $-$151	& 0.967 &	0.033 &	0.344 &	0.656 &	0.561 &	0.439 &	0.334 & 0.784 &	0.544 &	0.226 \\ 
4549	&	9290	& 9163	   & $-$126	& 0.955 &	0.045 &	0.348 &	0.652 &	0.597 &	0.403 &	0.362 &	0.791 & 0.544 &	0.242 \\
3871	&	7936	& 7835	   & $-$100	& 0.930 &	0.070 &	0.357 &	0.643 &	0.646 &	0.354 &	0.411 &	0.806 & 0.541 &	0.268 \\
3193	&	6582	& 6511	   & $-$71	& 0.885 &	0.115 &	0.372 &	0.628 &	0.730 &	0.270 &	0.475 &	0.833 & 0.536 &	0.301 \\
2515	&	5230	& 5189	   & $-$41	& 0.778 &	0.222 &	0.407 &	0.593 &	0.795 &	0.205 &	0.621 &	0.919 & 0.523 &	0.369 \\
1836	&	3878	& 3865	   & $-$13	& 0.579 &	0.421 &	0.474 &	0.526 &	0.880 &	0.120 &	0.966 &	1.181 & 0.499 &	0.530 \\
1158	&	2534	& 2552	   & $>$ 0 & 0.333 &	0.667 &	0.556 &	0.444 &	1.000 &	0.000 &$\gg$ 1&$\gg$ 1& 0.470 &	$\gg$ 1 \\
  \end{tabular}\end{ruledtabular}
	\label{Tab2}
\end{table}

The meson masses of the threshold in this model  are given in Table~\ref{Tab1}, together with the masses 
of other mesons that will be involved in the weak decays discussed in Sec.~\ref{secIII}. Also 
shown is the quark-antiquark r.m.s.\ radius. 

One can now study the stability of the $J^P=1^+$ $T^-_{bb}$ isoscalar state. In the GGV method, if 
a state is  unbound, one observes a slow decrease of its mass toward $M_1 + M_2$ 
as $N$, the number of terms  in Eq.~(\ref{Rkr}), increases. It turns out to be useful to also 
look at the content of the variational wave  function, which comes very close 
to 100\% in a color singlet-singlet channel in the physical basis~\cite{Via09}. On the other hand, if a 
variational state converges to a bound state as $N$ increases,  it includes sizable hidden-color 
components even for low $N$. We show in Table~\ref{Tab2} the results for the $T_{QQ}$ tetraquark 
for different masses of the heavy quark, $M_Q$. In the first line we give the results for 
the standard mass value of the bottom quark used in the
AL1 model, for which we get a binding energy of 151\,MeV. We have scrutinized the structure
of the $T_{QQ}$ state. For each particular value of $M_Q$ we have evaluated the lowest
strong-decay threshold, $M_1 + M_2$, the energy of the four-quark state, $M_{T_{QQ}}$, and
the corresponding binding energy $B=-\Delta E$. We have calculated the probability of the
$\bar 3 3$, $P[| \bar 3 3\rangle]$, and $6 \bar 6$,
$P[| 6 \bar 6\rangle]$, color components. By using the recoupling techniques
derived in Ref.~\cite{Via09} we have also evaluated the probability of the $1 1$,
$P[\ka]$, and $8 8$, $P[\kb]$, color components. Afterwards, we have 
expanded the wave function in terms of physical states evaluating the
probability of the pseudoscalar-vector, $P_{MM^*}$, and vector-vector, $P_{M^*M^*}$,
two-meson physical states. Finally we have calculated the average distance between the two heavy
quarks, $\langle x^2\rangle^{1/2}$, between the two light quarks, $\langle y^2\rangle^{1/2}$,
between a heavy and a light quarks, $\langle z^2\rangle^{1/2}$, 
and the four-quark r.m.s. radius, $X_{T_{QQ}}$.
\begin{figure}[t]
\centering
\hspace*{-0.5cm} 
\includegraphics[width=.5\columnwidth]{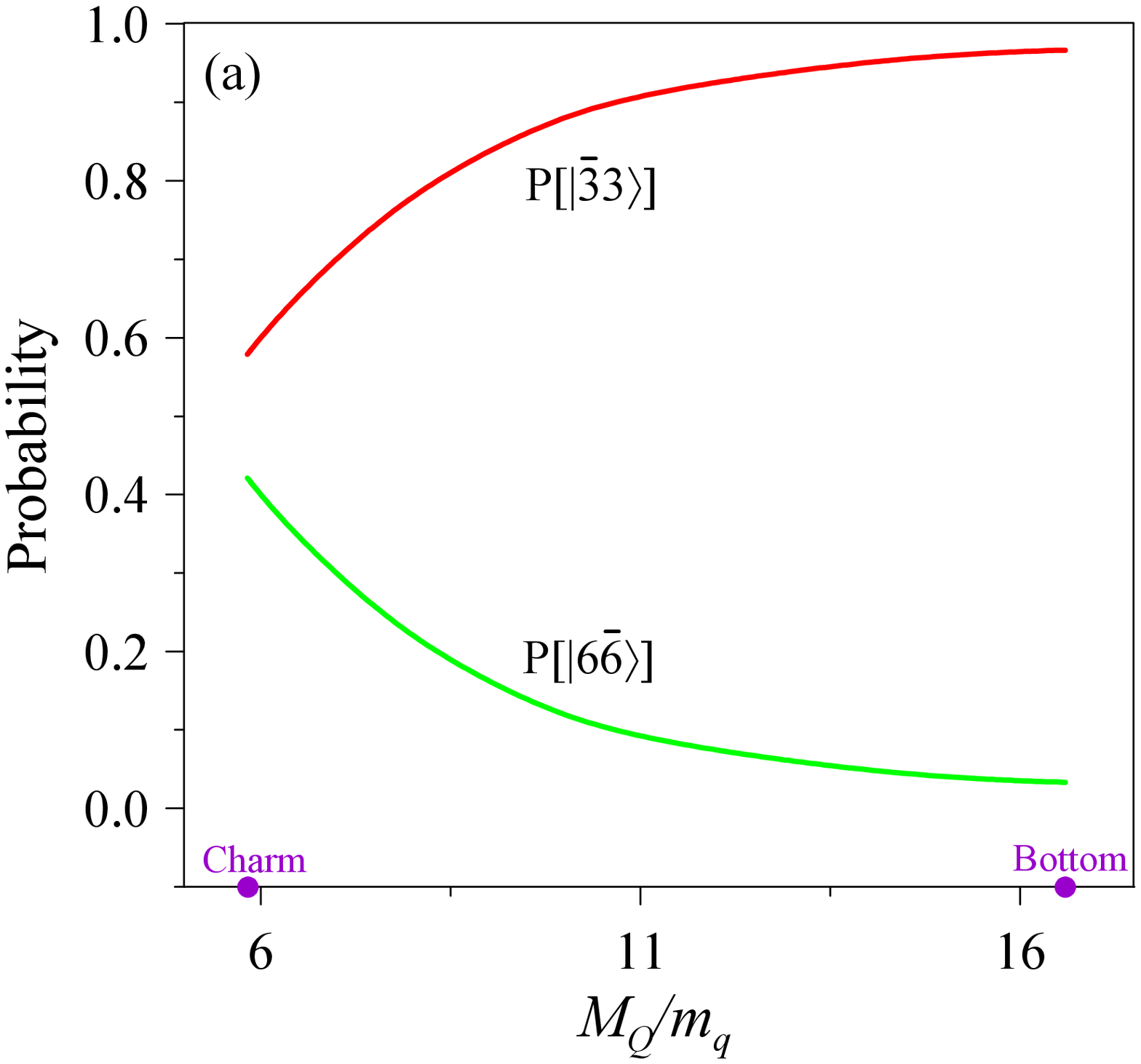}
\includegraphics[width=.5\columnwidth]{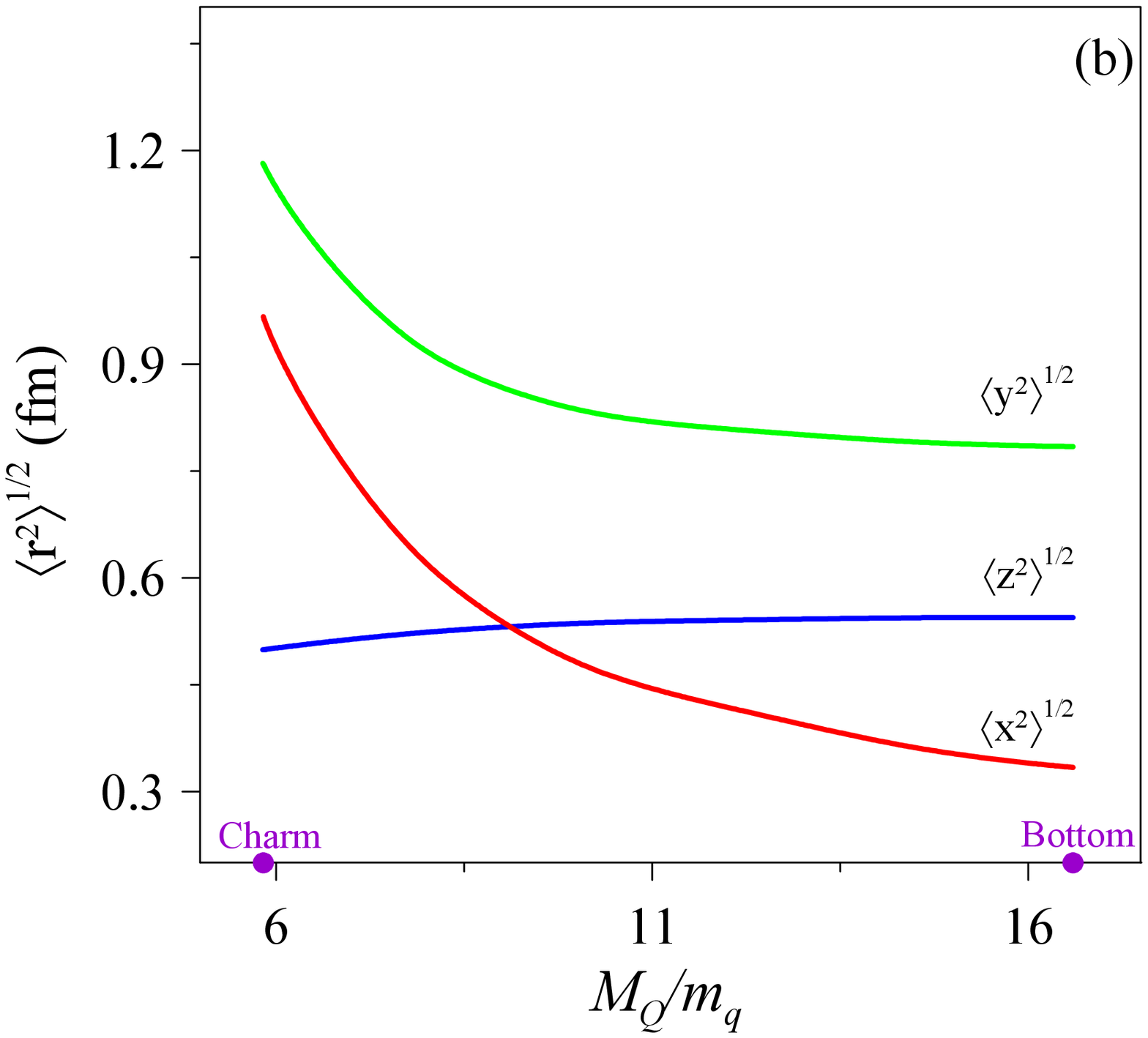}
\vspace*{-5.5cm}
\caption{(a) Probability of the $\bar 3 3$ and $6 \bar 6$ components
of the $T_{QQ}$ color wave function as a function of $M_Q/m_q$.
(b) Average distances $\langle \rm{x}^2\rangle^{1/2}$, $\langle \rm{y}^2\rangle^{1/2}$ 
and $\langle \rm{z}^2\rangle^{1/2}$ as a function of $M_Q/m_q$.}
\label{Fig1}
\end{figure}

As shown in Table~\ref{Tab2} the binding energy of the $T_{QQ}$ tetraquark increases
with increasing $M_Q/m_q$ as predicted in Ref.~\cite{Ade82} and recently rediscovered in 
Refs.~\cite{Eic17,Kar17}. Close to $\Delta E = 0$ the system behaves like
a simple meson-meson molecule, with a large probability in a single meson-meson
component, the pseudoscalar-vector channel. 
The $T_{QQ}$ starts to be bound around the mass of the charm
quark used by the AL1 model, $m_c=$ 1836\,MeV~\cite{Jan04}.
Such small binding is due to a cooperative effect between the 
chromoelectric and chromomagnetic pieces in the interacting potential. 
Hence, the $T_{cc}$ tetraquark is unbound when the spin-spin 
interaction is switched off. The $\llcol{i}{j}$ contribution 
of Eq.~(\ref{eq:mod1}), with a pairwise potential due to color-octet exchange, induces 
mixing between $\bar33$ and $6\bar6$ color states in the $QQ-\bar q\bar q$ basis.
The ground state of the $QQ\bar u\bar d$ with $J^P=1^+$ has its dominant 
component with color $\bar33$, and spin $\{1,0\}$ in the $QQ-\bar u\bar d$ basis. The main admixture 
consists of $6\bar 6$ with spin $\{0,1\}$ and a symmetric orbital wave function.
Thus, for $M_Q/m_q$ close to the charm sector, the binding requires both the color 
mixing of $\bar33$ with $6\bar 6$, and the spin-spin interaction~\cite{Ric18,Bal83}. 
In the most advanced calculations of Ref.~\cite{Bal83}, it was acknowledged that a pure 
additive interaction will not bind $cc\bar q\bar q$, on the sole basis that 
this tetraquark configuration benefits from the strong $cc$ chromoelectric attraction 
that is absent in the $Q\bar q+Q\bar q$ threshold. In the case where 
$\bar q \bar q= \bar u \bar d$, however, there is in addition a favorable 
chromomagnetic interaction in the tetraquark, while the threshold experiences 
only heavy-light spin-spin interaction, whose strength is suppressed by a factor $m_q/M_Q$. 

When the ratio $M_Q/m_q$ increases, the probability of the $6 \bar 6$ color component
diminishes in such a way that the system does not behave any more like a simple meson-meson 
molecule. The probability of the $6 \bar 6$ component in a compact $QQ\bar q\bar q$ tetraquark tends 
to zero for $M_Q \to \infty$. Therefore, heavy-light compact bound states would be almost a pure $\bar 3 3$ 
singlet color state and not a single colorless meson-meson $ 1 1$ molecule, as shown in Table~\ref{Tab2}.
Such compact states with two-body colored components can be expanded as the mixture of several 
physical meson-meson channels~\cite{Har81,Vij09}, and thus they can be also studied as an involved
coupled-channel problem of physical meson-meson states~\cite{Ike14}.
\begin{figure}[t]
\centering
\vspace*{-1.0cm} 
\includegraphics[width=0.90\columnwidth]{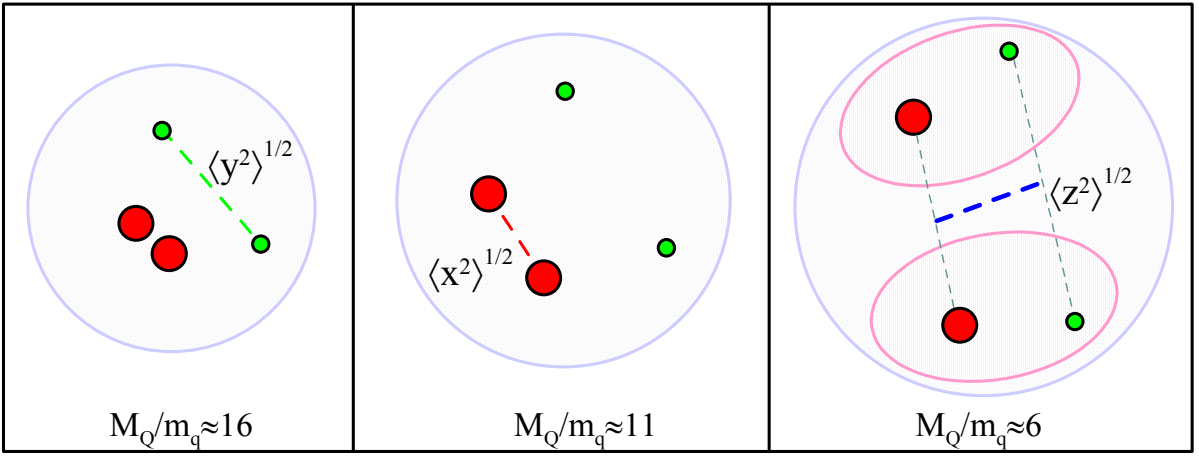}
\vspace*{-15.0cm}
\caption{From  left to  right, schematic evolution of a $T_{QQ}$ state as the heavy-quark mass decreases and, thus, the 
separation between the heavy quarks increases. The separation between the light quarks starts to augment close to 
threshold and the separation between the heavy and the light quarks remains almost constant. The last scenario,
$M_Q/m_q \approx 6$ is close to threshold, compatible with a $T_{QQ}$ molecule or two heavy-light mesons.}
\label{Fig2}
\end{figure}

We have shown these results in Fig.~\ref{Fig1}.
In the panel (a),  we see how the probability of the $6 \bar 6$ color component tends to
zero for $M_Q \to \infty$. On the other hand, we can also see the failure of treating
heavy-light tetraquarks as a single $\bar33$ color state for charm-light or charm-strange doubly-heavy tetraquarks.
In the panel (b) of Fig.~\ref{Fig1} we show the expectation value
of the different Jacobi coordinates over the tetraquark wave function, i.e., the average 
distance between the different constituents of the tetraquark~\cite{Vij09}. 
One can see how when the binding increases, i.e. $M_Q/m_q$
augments, the average distance between the two heavy quarks, $\langle {x}^2\rangle^{1/2}$, diminishes
rapidly, while that of the two light quarks, $\langle {y}^2\rangle^{1/2}$, although diminishing,
remains larger. The heavy-to-light quark distance, $\langle {z}^2\rangle^{1/2}$, stays almost 
constant for any value of $M_Q/m_q$. It is also worth noting how the tetraquark becomes compact 
in the bottom sector. As can be seen from Tables~\ref{Tab1} and ~\ref{Tab2}, for deep binding,
$X_{T_{QQ}}/{\rm{r.m.s.}}(M_1 + M_2)=0.226/0.298 < 1$, the tetraquark is smaller than the two mesons
of the threshold while close to $\Delta E = 0$, $X_{T_{QQ}}/{\rm{r.m.s.}}(M_1 + M_2)=0.530/0.464 > 1$,
it becomes larger, being very likely the break down into two mesons.
Thus, in the heavy-quark limit, the lowest lying tetraquark configuration
resembles the Helium atom~\cite{Lip86,Eic17}, a factorized system with separate dynamics
for the compact color $\bar 3$ $QQ$ {\it kernel} and for the
light quarks bound to the stationary color $3$ state, to construct
a $QQ\bar q \bar q$ color singlet. As mentioned above, this result is less pronounced for other
systems like charm-light ($cc\bar q\bar q$) or charm-strange ($cs\bar q\bar q$) doubly-heavy 
tetraquarks. On the basis of the results shown in Table~\ref{Tab2},
the schematic evolution of the $T_{QQ}$ state as a function of the ratio $M_Q/m_q$,
in other words, from deep binding to a close-to-threshold meson-meson state, 
is shown in Fig.~\ref{Fig2}~\cite{Qui18}. 

\section{Tetraquark lifetime and decay modes.} 
\label{secIII}
The double beauty $T^-_{bb}$ isoscalar tetraquark  with   $J^P=1^+$ is 
stable with respect to strong- and electromagnetic 
interactions~\cite{Ade82,Eic17,Kar17,Fra17,Bic16,Vij09,Ric18,Luo17,Duc13,Cza18}, 
and thus it decays weakly.  
We have studied the semileptonic and nonleptonic decays of  $T^-_{bb}$
following closely the method developed in Ref.~\cite{Her06}.
We present here the results for the most favorable final states where  
$T^-_{bb}$  might be looked for. The remaining channels and a detailed discussion
of the technicalities will be presented elsewhere~\cite{Her19}.

\begin{figure}[t]
\centering
\includegraphics[width=0.85\columnwidth]{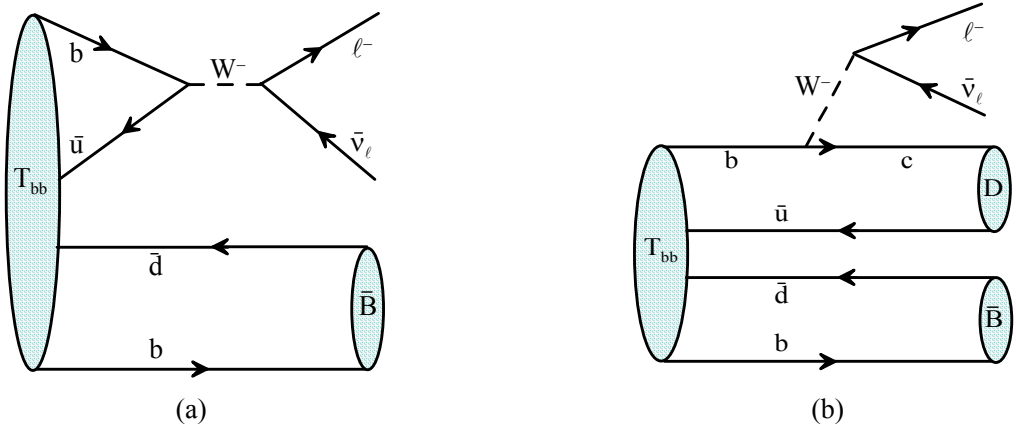}
\vspace*{-16.0cm}
\caption{Representative diagrams for semileptonic decays of the $T_{bb}$ 
tetraquark: (a) Final state with a single meson. (b) Final state with two mesons.}
\label{Fig3}
\end{figure}

Among the semileptonic decays one can distinguish between processes with final states with a single meson, see
panel (a) of Fig.~\ref{Fig3}, or those with two mesons, panel (b) of Fig.~\ref{Fig3}. The first case,
$T^-_{bb}\to \bar B^{0} \,  \ell^- \, \bar\nu_\ell$, involves a $ b \bar u \to W^- \to \ell^- \, \bar\nu_\ell$
transition that at tree level is described by the operator
\be
-iV_{ub}\frac{G_F}{\sqrt2
}\, \overline\Psi_u(0)\gamma^\mu(1-\gamma_5)\Psi_b(0)\,
\overline{\Psi}_\ell(0)\gamma_\mu(1-\gamma_5)\Psi_{\nu_\ell}(0) \, ,
\ee
where $\Psi_f$ is a quark field of a definite flavor $f$, $G_F$ is the Fermi coupling 
constant and $V_{ub}$ is the Cabibbo-Kobayashi-Maskawa (CKM) 
matrix element. The decay width is given by
\begin{multline}
\Gamma =\frac1{2m_T}\iiint \frac{d \vec P_B}{(2\pi)^32E_B}
\frac{d \vec p_\ell}{(2\pi)^32E_\ell}
\frac{d \vec p_{\nu_\ell}}{(2\pi)^32E_{\nu_\ell}}\,(2\pi)^4
\delta^{(4)}(P_T-P_B-p_\ell-p_{\nu_\ell}) \\ 
{}\times \, |V_{ub}|^2\frac{G_F^2}{2}{\cal
L}^{\alpha\beta}(p_\ell,p_{\nu_\ell})\,{\cal W}_{\alpha\beta}(P_T,P_B) \, , \label{gg}
\end{multline}
where the lepton\footnote{The $\pm$ signs correspond respectively to decays into $\ell^- \bar \nu_\ell$ 
and $\ell^+ \nu_\ell$.} and hadron tensors are given by,
\begin{align}
{\cal L}^{\alpha\beta} &=
8( p_\ell^\alpha p_{\nu_\ell}^\beta + 
   p_\ell^\beta  p_{\nu_\ell}^\alpha -
   g^{\alpha\beta} p_\ell\cdot p_{\nu_\ell}
\pm i\epsilon^{\alpha\beta\rho\lambda}p_{l\rho}p_{{\nu_\ell}\lambda}) \, ,\\
\mathcal{W}_{\alpha\beta}&=\frac1{2J_T+1}\sum_{\lambda ,\lambda^\prime}h^{T\to B}_\alpha\,(h^{T\to B}_\beta)^*\,\\
h^{T\to B}_\alpha&=\left\langle {B,\lambda^\prime\,\vec P_B}
\right|\overline\Psi_u(0)\gamma_\alpha(1-\gamma_5)\Psi_b(0)\left| {T,\lambda\,\vec{0}}
\right\rangle\,,\label{eq:hadr-cur}
\end{align}
\noindent
where $p_i$ is the four-momentum of the particle $i$, $J_T$ stands for the
spin of the  tetraquark, $|{M,\lambda^\prime \,\vec P_M}\rangle$ represents the
state of an $M$ meson  with three momentum $\vec P_M$ and 
spin projection in the meson center of mass $\lambda^\prime$,
and $|{T,\lambda\,\vec{0}}\,\rangle$ is the state
of the  tetraquark at rest. $\epsilon^{\alpha\beta\rho\lambda}$
is the fully antisymmetric tensor for which we take the convention
$\epsilon^{0123}=+1$ and $g^{\alpha\alpha}=(1,-1,-1,-1)$.
Equation~(\ref{gg}) can be further simplified
\be
\Gamma=|V_{ub}|^2\frac{G_F^2}{2^{4}\pi^3 m_{T}}\iint dE_B dE_\ell\,\Theta(m_{T}-E_B-E_\ell)\,
\Theta(1-|\cos\theta_\ell^0|)\ 
\tilde{\cal L}^{\alpha\beta}\,{\cal W}_{\alpha\beta}(P_{T},\tilde P_B) \, ,
\label{eq:semi_I}
\ee
with $\tilde P_B^\mu=(E_B,0,0,|\vec{P}_B|)$ and 
\be
\cos\theta_\ell^0=\frac{(m_{T}-E_B-E_\ell)^2-|\vec P_B|^2-|\vec p_\ell|^2}{2|\vec P_B|\,|\vec p_\ell|} \, .
\ee
In~\eqref{eq:semi_I}, since all the dependence on $\varphi_\ell$ appears
only in the lepton tensor, we have defined
\be
\tilde{\cal L}^{\alpha\beta}=\frac1{16\pi}\int d\varphi_\ell\,{\cal L}^{\alpha\beta} \, .
\ee
The matrix element~\eqref{eq:hadr-cur} appearing in the
hadron tensor can be expanded as,
\begin{multline}
h^{T\to B}_\rho = 4\,\sqrt{m_TE_B}
  \iint d \vec p_x d \vec p_z \sum_{\substack{\alpha_1,\alpha_2\\ \alpha_3,\alpha_4}}
\left[\hat{\phi}^{(B,\lambda')}_{\alpha_2,\alpha_4}
\left(\,\frac{m_b}{m_b+m_u}\,\vec{P}_B+\vec p_x-\frac12\,\vec p_z\right)\right]^*
\hat{\phi}^{(T,\lambda)}_{\substack{\alpha_1,\alpha_2 \\ \alpha_3,\alpha_4}} 
\left(\,\vec{p}_x,-\vec p_x-\vec{P}_B,\vec{p}_z\right) \\ 
{} \times \,
\frac{(-1)^{1/2-s_3}}{2\sqrt{E_uE_{b}}}\,
\bar v^{s_3}_u\Big(-\vec{P}_B-\vec
p_x-\frac12\,\vec p_z\Big)\gamma_\rho(1-\gamma_5)
u_b^{s_1}\Big(\vec p_x+\frac12\vec
p_z\Big)\,\delta_{c_1c_3}\,\delta_{f_1 b}\,\delta_{f_3u} \, ,
\end{multline}
where $\hat{\phi}$ is the Fourier transform of the radial  wave function,  
obtained in Sec.~\ref{secII} using the AL1 constituent model, and 
$\alpha_i$ represents the quantum numbers of spin $s_i$, flavor $f_i$ and
color $c_i$ ($\alpha_i\equiv (s_i,f_i,c_i))$ of a quark or an antiquark. 

For the second class of semileptonic decays represented diagrammatically in panel (b) 
of Fig.~\ref{Fig3}, with a $b \to c$ transition at the quark level, the operator is given by
\be
-iV_{cb}\frac{G_F}{\sqrt2
}\, \overline\Psi_c(0)\gamma^\mu(1-\gamma_5)\Psi_b(0)\,
\overline{\Psi}_\ell(0)\gamma_\mu(1-\gamma_5)\Psi_{\nu_\ell}(0) \, ,
\ee
and the decay width can be expressed as
\begin{multline}
\Gamma = |V_{cb}|^2\frac{G_F^2}{2^{7}\pi^5m_T}\iiiint|\vec{P}_B|dE_B
 d\cos\theta_D|\vec{P}_D|\,d{E_D}\frac1{|\vec{\tilde P}_B+\vec{P}_D|}
dE_\ell\,\Theta(1-|\cos\theta_\ell^0|) \\ 
{}\times \, \Theta(m_T-E_B-E_D-E_\ell)\ \tilde{\cal
L}^{\alpha\beta}\,{\cal W}_{\alpha\beta}(P_{T},\hat P_B,\hat P_D) \, ,
\end{multline}
where $\hat P_B = \mathcal{R}^\prime \tilde P_B$ and $\hat P_D = \mathcal{R}^\prime P_D$, where  $\mathcal{R}^\prime$
is a rotation that, for a fixed $\vec P_D$, takes $\vec{\tilde P}_B+\vec{P}_D
\to(0,0,|\vec{\tilde P}_B+\vec{P}_D|)$. In this case,
\be
\cos\theta_\ell^0=\frac{(m_{T}-E_B-E_D-E_\ell)^2-|\vec{\tilde P}_B+\vec{P}_D|^2-|\vec p_\ell|^2}
{2|\vec{\tilde P}_B+\vec{P}_D|\,|\vec p_\ell|} \, .
\ee

The matrix element appearing in the hadron tensor
\be 
h^{T\to M_1M_2}_\rho = \left\langle {M_1,\lambda^{\prime \prime} \,\vec{P}_1}\right|\left\langle {M_2,\lambda^\prime\,\vec{P}_2}
\right|\overline\Psi_c(0)\gamma_\rho(1-\gamma_5)\Psi_b(0)\left|{T,\lambda\,\vec{0}}\,\right\rangle \, ,
\label{hTbb}
\ee
here written for a $T \to BD$ transition (for other cases, the  changes are obvious),
can be expressed as,
\be
\begin{aligned}
h^{T\to BD}_\rho =4\,(2\pi)^{3/2}&\sqrt{2m_TE_BE_D}
 \sum_{\substack{\alpha_2,\alpha_3 \\ \alpha_4, \alpha_5}}
 \iint d \vec p_x   d \vec p_z \left[\hat{\phi}^{(D,\lambda^{\prime\prime})}_{\alpha_5,\alpha_3}
\left(\,-\frac{m_u}{m_c+m_u}\vec{P}_D-\vec{P}_B -\vec
p_x-\frac12\,\vec p_z\right)\right]^* \\
&\times\, \left[\hat{\phi}^{(B,\lambda^\prime)}_{\alpha_2,\alpha_4}
\left(\,\frac{m_b}{m_b+m_u}\,\vec{P}_B+\vec p_x-\frac12\,\vec p_z \right)\right]^*
\sum_{\alpha_1}
\hat{\phi}^{(T,\lambda)}_{\substack{\alpha_1,\alpha_2 \\ \alpha_3,\alpha_4}}
\left(\,\vec p_x,-\vec p_x-\vec{P}_B,\vec{p}_z \right) \\
&\times \, \frac{1}{2\sqrt{E_cE_{b}}}\,
\bar u^{s_5}_c\Big(\vec{P}_D+\vec{P}_B+\vec
p_x+\frac12\,\vec p_z\Big)\gamma_\rho(1-\gamma_5)
u_b^{s_1}\Big(\vec p_x+\frac12\vec
p_z\Big)\,\delta_{c_1c_5}\delta_{f_1b}\,\delta_{f_5c} \, .
\end{aligned}\ee
Obviously $B$ could also be a $B^*$ and $D$ a $D^*$.
If we have a $b\to u $ quark transition, one has to change $V_{cb}\to
V_{ub}$ and the meson in the final state (apart from $B(B^*)$) would be 
a nonstrange meson with $u\bar u$ or $u\bar d$ composition. 
\begin{figure}[t]
\centering
\includegraphics[width=0.85\columnwidth]{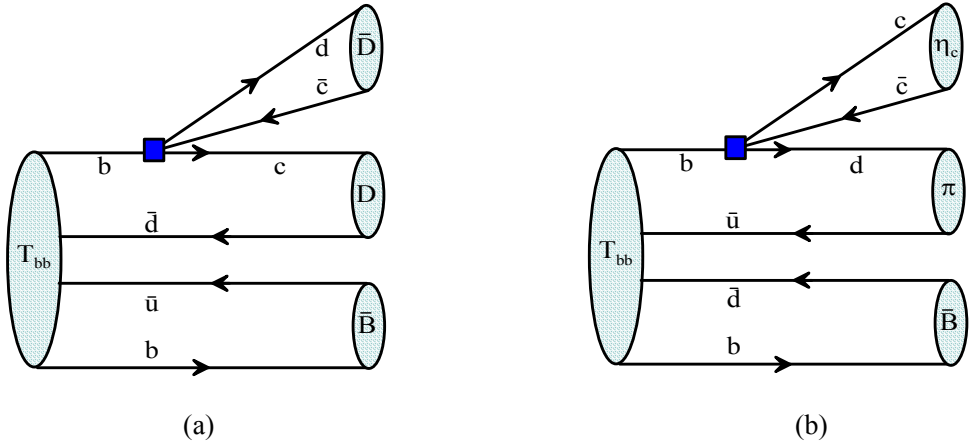}
\vspace*{-16.0cm}
\caption{Representative diagrams for nonleptonic decays of the $T_{bb}$ tetraquark:
(a) Final state with only open flavor mesons. (b) Final state with hidden flavor mesons. The
blue box represents a four-quark effective vertex containing the contribution of
the $W$ boson and radiative corrections as seen in Eq.~\eqref{heff}.}
\label{Fig4}
\end{figure}

We evaluate now the width of the nonleptonic decays
$T^-_{bb}\to \, B^- \, M_1 \, M_2$ or $ \bar B^0 \,M^\prime_1 \, M^\prime_2$
represented diagrammatically in Fig.~\ref{Fig4}. These decay modes involve 
a transition $b\to c,u$ at the quark level and they are governed, neglecting 
penguin operators, by the effective Hamiltonian~\cite{Ebe03,Iva06,Col00}
\be
H_\text{eff}=\frac{G_F}{\sqrt2}\,\left\{ V_{cb}\left[
c_1(\mu)\,Q_1^{cb}+c_2(\mu)\,Q_2^{cb}\right]+V_{ub}\left[
c_1(\mu)\,Q_1^{ub}+c_2(\mu)\,Q_2^{ub}
\right] + \text{h.c.}\right\} \, ,
\label{heff}
\ee
where $c_1,\,c_2$ are scale--dependent Wilson coefficients, and
$Q_1^{ib},\,Q_2^{ib}$, $i=u,c$, are local four-quark 
operators of the current-current type given by
\begin{align}
&\begin{aligned}Q_1^{ib}=\overline{\Psi}_i(0)\gamma_\mu(1-\gamma_5)\Psi_b(0)\,\bigg[&
\ V_{ud}^*\, \overline{\Psi}_d(0)\gamma^\mu(1-\gamma_5)\Psi_u(0)+
V_{us}^*\, \overline{\Psi}_s(0)\gamma^\mu(1-\gamma_5)\Psi_u(0)\\
& {}+
V_{cd}^*\, \overline{\Psi}_d(0)\gamma^\mu(1-\gamma_5)\Psi_c(0)+
V_{cs}^*\, \overline{\Psi}_s(0)\gamma^\mu(1-\gamma_5)\Psi_c(0)\,
\bigg] \, ,\
\end{aligned}
\\[2pt]
&\begin{aligned} Q_2^{ib}{}={}&\overline{\Psi}_d(0)\gamma_\mu(1-\gamma_5)\Psi_b(0)\,\bigg[
\ V_{ud}^*\, 
\overline{\Psi}_i(0)\gamma^\mu(1-\gamma_5)\Psi_u(0)
+ V_{cd}^*\, \overline{\Psi}_i(0)\gamma^\mu(1-\gamma_5)\Psi_c(0)\,
\bigg]\\
{}+{}&\overline{\Psi}_s(0)\gamma_\mu(1-\gamma_5)\Psi_b(0)\, \bigg[
V_{us}^*\, \overline{\Psi}_i(0)\gamma^\mu(1-\gamma_5)\Psi_u(0)\,
+V_{cs}^*\,\overline{\Psi}_i(0)\gamma^\mu(1-\gamma_5)\Psi_c(0)
\bigg] \, ,
\end{aligned}
\end{align}
where the different $V_{jk}$ are CKM matrix elements. 
\begin{table}[b]
\caption{Meson decay constants, in GeV, used in this work.}
\begin{ruledtabular}
  \begin{tabular}{ccccccc}
	$f_{\pi^-}$ & $f_{\pi^0}$ & $f_{\rho^-,\rho^0}$ & $f_{D^+}$ & $f_{{D^*}^+}$ & $f_{D_s^+}$ & $f_{{D_s^*}^+}$ \\
\hline \\[-9pt]
0.1307~\cite{Pdg18} &
0.130~\cite{Pdg18}  & 
0.210~\cite{Iva06}  &
0.2226~\cite{Art05} & 
0.245~\cite{Bec99}  &
0.294~\cite{Pdg18}  &
0.272~\cite{Bec99} 
  \end{tabular}
	\label{Tab3}
\end{ruledtabular}	
\end{table}

We work in  the factorization approximation which amounts to evaluate 
the hadron matrix elements of the effective Hamiltonian as a product of two
quark-current matrix elements: one is the matrix element for the 
$T_{bb} \to B M_1$ transition, and the other accounts for  the transition
from vacuum to the other final meson $M_2$, see Fig~\ref{Fig4}.
The latter coupling is governed by the corresponding meson decay constant.
When writing the factorization amplitude, the relevant coefficients of 
the effective Hamiltonian~\eqref{heff} are the combinations,
\begin{equation}
a_1(\mu) = c_1(\mu)+\frac{1}{N_C}\,c_2(\mu)\hspace{0.5cm} \qquad
a_2(\mu)= c_2(\mu)+\frac{1}{N_C}\,c_1(\mu) \, ,
\end{equation}
with $N_C=3$ the number of colors. The energy scale $\mu$ appropriate in this
case is $\mu\simeq m_b$ and the values for $a_1$ and $a_2$ that we use
are~\cite{Iva06}:
\begin{equation}
a_1 = 1.14\qquad
a_2 = -0.20 \, .
\end{equation}
Note that the $W$-exchange diagrams, that play an important role in the decay of charm, are 
suppressed in the decay of $b$ since they are proportional to $a_2$. 
The total decay width is given as
\begin{multline}
\Gamma = \frac1{2m_{T}}\iiint \frac{d \vec P_B}{(2\pi)^32 E_B}
\frac{d \vec P_1}{(2\pi)^32E_1}  \frac{d \vec P_2}{(2\pi)^32E_2}
\,(2\pi)^4 \delta^{(4)}(P_{T}-P_B-P_1-P_2)  \\ 
{}\times \,
 \frac{G_F^2}{2}\frac1{2J_T+1}
 \sum_{\substack{\lambda_T, \lambda_B \\ \lambda_1, \lambda_2}}|{\cal
 M}_{\lambda_T \lambda_B \lambda_1 \lambda_2}(P_{T},P_B,P_1,P_2)|^2 \, .
\end{multline}
Using invariance arguments as in the semileptonic decay case 
one finds,
\begin{multline}
\Gamma = \frac{G_F^2}{2^{7}\pi^3m_T}
\iint dE_B d{E_1}\ \Theta(1-|\cos\theta_1^0|)\,\Theta(m_T-E_B-E_1-M_2) \\ 
{}\times \, \frac1{2J_T+1}\sum_{\substack{\lambda_T, \lambda_B \\
\lambda_1, \lambda_2}} \left| {\cal
M}_{\lambda_T\lambda_B\lambda_1\lambda_2}(P_T,\tilde P_B,\hat P_1,\hat P_2) \right|^2 \, ,
\end{multline}
where 
\be
\cos\theta_1^0=\frac{(m_{T}-E_B-E_1)^2-M_2^2-|\vec{ P}_B|^2-|\vec{P}_1|^2}
{2\,|\vec{ P}_B|\,|\vec{P}_1|} \, ,
\ee
and ${\cal M}$ involves the product of a hadron matrix element such as Eq.~\eqref{hTbb} and
meson decay constants that are taken from experiment or lattice data. 
For instance, for a $T_{bb}^-\to B^-D^+D^-$ decay, one has that
\be
{\cal
M}=V_{cb}V_{cd}^*\,a_1\,h^{T\to B^-D^+}_\alpha if_{D^-}P^\alpha_{D^-},
\ee
In particular, for the  decays  presented in Table~\ref{Tab6}, we have used
the meson decay constants listed in Table~\ref{Tab3}.\\

\begin{table}[b]
\caption{Decay widths, in units of $10^{-15}$\,GeV, for processes described by Fig.~\ref{Fig3}(a).}
\begin{tabular}{lp{1cm}c}\hline\hline
	Final state & & $\Gamma \, [10^{-15}\,\rm{GeV}]$ \\
	\hline
 $ \bar {B^*}^{0} \, e^- \, \bar \nu_e $           & & $0{.}0365 \pm 0{.}0004$\\
 $ \bar B^{0} \, e^- \, \bar \nu_e $               & & $0{.}0394 \pm 0{.}0006$\\
 $ \bar {B^*}^{0} \, \mu^- \, \bar \nu_{\mu} $     & & $0{.}0355 \pm 0{.}0004 $\\
 $ \bar B^{0} \, \mu^- \, \bar \nu_{\mu} $         & & $0{.}0396 \pm 0{.}0006 $\\
 $ \bar {B^*}^{0} \, \tau^- \, \bar \nu_{\tau} $   & & $0{.}0355 \pm 0{.}0004 $\\
 $ \bar B^{0} \,  \tau^- \, \bar \nu_{\tau} $      & & $0{.}0396 \pm 0{.}0006 $\\\hline\hline
  \end{tabular}
	\label{Tab4}
\end{table}

\begin{figure}[t]
\centering
\includegraphics[width=0.85\columnwidth]{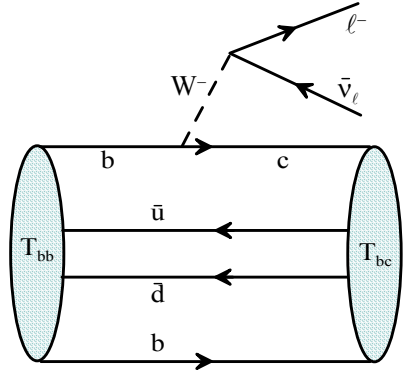}
\vspace*{-16.0cm}
\caption{Representative diagram for the semileptonic decay of the $T_{bb}$ $J^P=1^+$ 
tetraquark to the $T_{bc}$ $J^P=0^+$ tetraquark.}
\label{Fig5}
\end{figure}

For the sake of completeness we have also evaluated the decay of the $J^P=1^+$ $T_{bb}^-$ 
isoscalar tetraquark into the $J^P=0^+$ $T_{bc}^0$ isoscalar tetraquark, decay depicted in 
Fig.~\ref{Fig5}. The mass of the  $J^P=0^+$ $bc\bar u\bar d$ isoscalar state
has been estimated in  Ref.~\cite{Kar17} where the authors
obtain a central value 11\,MeV below the $\bar B D$ threshold, although it is cautioned 
that the precision of the calculation is not sufficient to determine whether the 
tetraquark is actually above or below this threshold.
A systematic study of exotic $QQ^\prime \bar q\bar q$ four-quark states containing distinguishable 
heavy flavors, $b$ and $c$, has been recently performed with the AL1 model in Ref.~\cite{Car19}. 
The $J^P=0^+$ isoscalar state was found to be strong and electromagnetic-interaction stable with a binding
energy of around 23 MeV. Other independent calculations made in different frameworks 
arrive to similar conclusions. Among them, it is important to emphasize 
 the lattice QCD results of Ref.~\cite{Fra19} 
where it is found evidence for the existence of a strong-interaction-stable $(I)J^P=(0)1^+$ 
$bc\bar u\bar d$ four-quark state with a mass in the range of 15 to 61 MeV below 
the $D\bar B{}^*$ threshold. The decay width in this case is given by~\eqref{eq:semi_I}, 
changing the final $B$ meson by the $T_{bc}^0$ tetraquark
and $V_{ub}$ by $V_{cb}$, while the corresponding hadronic matrix element is 
\be
\begin{aligned}
h^{T\to T_{bc}}_\rho =2\,&\sqrt{2m_TE_{T_{bc}}}
 \sum_{\substack{\alpha_2,\alpha_3 \\ \alpha_4, \alpha_5}}
 \iiint d \vec p_x  d \vec p_y d \vec p_z \\
 &\times\,\left[\hat{\phi}^{(T_{bc},\lambda^\prime)}_{\substack{\alpha_2,\alpha_5
 \\ \alpha_3,\alpha_4}}
\left(\,-\vec p_x-\frac{m_b-m_c}{2(m_b+m_c)}\,\vec p_z-
\frac{m_b}{m_b+m_c}\,\vec P_{T_{bc}},\vec p_y,\vec p_z+\frac{2m_u}{m_b+m_c+2
m_u}\vec{P}_{T_{bc}}\right)\right]^* \\
&\times\, 
\sum_{\alpha_1}
\hat{\phi}^{(T,\lambda)}_{\substack{\alpha_1,\alpha_2 \\ \alpha_3,\alpha_4}}
\left(\,\vec p_x,\vec p_y,\vec{p}_z \right) 
\, \frac{1}{2\sqrt{E_cE_{b}}}\,
\bar u^{s_5}_c\Big(\vec{P}_{T_{bc}}+\vec
p_x+\frac12\,\vec p_z\Big)\gamma_\rho(1-\gamma_5)
u_b^{s_1}\Big(\vec p_x+\frac12\vec
p_z\Big)\,\delta_{c_1c_5}\delta_{f_1b}\,\delta_{f_5c} \, .
\end{aligned}\ee

Let us now comment on the results. Some aspects could have been anticipated, 
and are verified. For instance, for the $T\to B^{(*)}$ semileptonic decays 
depicted in Fig.~\ref{Fig3}(a), and due to the large phase space available in
all cases, the differences among the widths  into the three lepton families 
are very small. The corresponding results\footnote{The errors 
quoted correspond to  the uncertainties of the Monte 
Carlo numerical integration.} are shown in Table~\ref{Tab4}.
We also note that the overlap in the hadron tensor between the $T$ and 
the $B (B^*)$ wave function slightly favors the pseudoscalar mesons.
Anyhow, decays with a single meson in the final state are suppressed 
by at least two orders of magnitude as compared to the semileptonic decays 
with two final  mesons, and the leading non-leptonic modes that are discussed below. 

For semileptonic decays involving two mesons in the final state, described by 
panel (b) of Fig.~\ref{Fig3}, the processes involving a $ b \to c$ vertex are 
favored compared to those involving a $b \to u$ vertex, due to the larger CKM 
matrix element $\left| V_{cb} \right|\sim 0{.}041$ compared to 
$\left| V_{ub} \right|\sim 0{.}0035$~\cite{Pdg18}. In Table~\ref{Tab5} we show
the most favorable channels, the filter being a width larger than 
$10^{9}\,{\rm s}^{-1}=0{.}66 \times 10^{-15}$\,GeV,
for the semileptonic decays with two mesons and a light $\ell=e,\mu$ lepton 
in the final state. Though much smaller, we also give the widths for the
corresponding channels with a final $\tau$ since they could be interesting in
the context of studies of lepton-flavor universality violation.
Due to spin recoupling coefficients, the largest decay widths appear for vector 
mesons in the final state. In short, the largest preferred semileptonic decay 
are $B^{(*)} \, D^{(*)} \, \ell \, \bar \nu_\ell$ with the
various combinations of spins for the mesons, and $\ell=e,\mu$. 
\begin{table}[t]
\caption{Largest decay widths, in units of $10^{-15}$\,GeV, for 
the processes described by Fig.~\ref{Fig3}(b). Here $\ell=e,\mu$.}
  \begin{tabular}{lp{1cm}cp{0.3cm}p{0.3cm}lp{1cm}c}\hline\hline
	Final state & &$\Gamma \, [10^{-15}\,\rm{GeV}]$ &&&Final state & &$\Gamma \, [10^{-15}\,\rm{GeV}]$\\
	\hline
 ${B^*}^{-} \, {D^*}^{+} \, \ell^- \, \bar \nu_\ell $                    &  & \multirow{2}{*}{$9{.}02 \pm 0{.}07 $}&&&
${B^*}^{-} \, {D^*}^{+} \, \tau^- \, \bar \nu_\tau $&&\multirow{2}{*}{$1.55\pm0.01 $}\\
 $\bar {B^*}^{0} \, {D^*}^{0} \, \ell^- \, \bar \nu_\ell $               &  &&&& $\bar {B^*}^{0} \, {D^*}^{0} \, \tau^- \, \bar \nu_\tau $\\
 ${B^*}^{-} \, D^{+} \, \ell^- \, \bar \nu_\ell $                        &  & \multirow{2}{*}{$3{.}59 \pm 0{.}03$}&&&
${B^*}^{-} \, D^{+} \, \tau^- \, \bar \nu_\tau$&&\multirow{2}{*}{$ 0.727\pm0.005$}\\
 $\bar {B^*}^{0} \, D^{0} \, \ell^- \, \bar \nu_\ell $                   &   &&&&  $\bar {B^*}^{0} \, D^{0} \, \tau^- \, \bar \nu_\tau $ \\
 $B^{-} \, {D^*}^{+} \, \ell^- \, \bar \nu_\ell $                        &  & \multirow{2}{*}{$4{.}63 \pm 0{.}05$}&&&
$B^{-} \, {D^*}^{+} \, \tau^- \, \bar \nu_\tau $&&\multirow{2}{*}{$ 0.86\pm0.007$} \\
 $\bar B^{0} \, {D^*}^{0} \, \ell^- \, \bar \nu_\ell $                   &  &&&& $\bar B^{0} \, {D^*}^{0} \, \tau^- \, \bar \nu_\tau $\\
 $B^{-} \,  D^{+} \, l^- \, \bar \nu_l $                           &  & \multirow{2}{*}{$1{.}92 \pm 0{.}02$}&&&
 $B^{-} \,  D^{+} \, \tau^- \, \bar \nu_\tau $&&\multirow{2}{*}{$ 0.409\pm0.003$}\\
 $\bar B^{0} \, D^{0} \,  \ell^- \, \bar \nu_\ell $                      &  &&&& $\bar B^{0} \, D^{0} \,  \tau^- \, \bar \nu_\tau $\\ \hline\hline
  \end{tabular}
	\label{Tab5}
\end{table}

Table~\ref{Tab6} displays now the most important 
nonleptonic decay modes. All of them contain a $b \to c$ vertex and an $a_1$
factor, and the dominant ones have a $D^{(*)}_{(s)}$ meson in the final
state. Once again  vector mesons are favored in the final state. 
As a consequence of the factorization approximation, processes with
$D_s$ or a light meson final states arising from vacuum have decay widths comparable to the 
corresponding semileptonic decay. This is due to the large value of the 
Cabibbo allowed CKM matrix elements 
$\left| V_{cs} \right|\sim \left| V_{ud} \right| \sim 0{.}97$~\cite{Pdg18} and the fact
that the hadronic matrix elements are proportional to 
$a_1^2$ in those cases. Decay channels not shown in Tables~\ref{Tab5} and~\ref{Tab6} 
are suppressed by at least one order of magnitude. For instance, final states 
with $J/\Psi$ or $\eta_c$ mesons, are suppressed by more than one order of magnitude 
since their widths are proportional to $|V_{cd}|^2a^2_2$. According to our study, the promising final 
states among the nonleptonic decays are
$\bar B^*{}^{-} \, D^*{}^{+} \, D_s^*{}^-$, $\bar B^*{}^{0} \, D^*{}^{0} \, D_s^*{}^- $, and 
$\bar B^*{}^{-} \, D^*{}^{+} \, \rho^-$.
\begin{table}[h!]
\caption{Largest decay widths, in units of $10^{-15}$\,GeV, for the
processes described by Fig.~\ref{Fig4}.}
  \begin{tabular}{lp{1cm}cp{2cm}lp{1cm}c}\hline\hline
	Final state & &$\Gamma \, [10^{-15}\,\rm{GeV}]$ & &Final state & &$\Gamma \, [10^{-15}\,\rm{GeV}]$ \\
	\hline
 ${B^*}^{-} \, {D^*}^{+} \, D_s^-$                               &   & \multirow{2}{*}{$4{.}00 \pm 0{.}06$} &&
 ${B^-} \, {D^*}^{+} \, {D_s^*}^- $                              &   & \multirow{2}{*}{$3{.}15 \pm 0{.}05$}\\
 $\bar {B^*}^{0} \, {D^*}^{0} \, D_s^- $                         &   & &&
 $\bar {B^0} \, {D^*}^0 \, {D_s^*}^- $                           &   & \\
 ${B^*}^{-} \, {D^*}^{+} \, {D_s^*}^-$                           &   & \multirow{2}{*}{$6{.}50 \pm 0{.}09$} &&
 $B^- \, D^+ \, {D_s^*}^- $                                      &   & \multirow{2}{*}{$1{.}20 \pm 0{.}02$}\\
 $\bar {B^*}^{0} \, {D^*}^{0} \, {D_s^*}^- $                     &   & &&
 $\bar {B^0} \, D^0 \, {D_s^*}^- $                               &   & \\
 ${B^*}^{-} \, D^{+} \, D_s^-$                                   &   & \multirow{2}{*}{$2{.}57 \pm 0{.}04$} &&
 ${B^*}^{-} \, {D^*}^{+} \, \rho^-$                              &   & $3{.}57 \pm 0{.}09$ \\
 $\bar {B^*}^{0} \, D^0 \, D_s^- $                               &   & &&
 ${B^*}^{-} \, {D^*}^{+} \, \pi^-$                               &   & $1{.}28 \pm 0{.}03$ \\
 ${B^*}^{-} \, D^{+} \, {D_s^*}^- $                              &   & \multirow{2}{*}{$2{.}32 \pm 0{.}03$} &&
 ${B^*}^{-} \, D^+ \, \rho^-$                                    &   & $1{.}70 \pm 0{.}04$ \\
 $\bar {B^*}^{0} \, D^0 \, {D_s^*}^- $                           &   & &&
 ${B^*}^{-} \, D^+ \, \pi^-$                                     &   & $0{.}70 \pm 0{.}02$ \\
 $B^- \, {D^*}^{+} \, D_s^-$                                     &   & \multirow{2}{*}{$2{.}78 \pm 0{.}05$} &&
 $B^- \, {D^*}^{+} \, \rho^-$                                    &   & $2{.}01 \pm 0{.}05$ \\ 
 $\bar {B^0} \, {D^*}^{0} \, D_s^- $                             &   & &&
 $B^- \, {D^*}^{+} \, \pi^-$                                     &   & $0{.}77 \pm 0{.}03$ \\\hline\hline
  \end{tabular}
	\label{Tab6}
\end{table}

Finally, in Table~\ref{Tab7} we show the results for the semileptonic decay
corresponding to Fig.~\ref{Fig5} with a $J^P=0^+$ $T_{bc}$ isoscalar tetraquark in the
final state. In our calculation, the total semileptonic decay width with a final 
$J^P=0^+$ $T_{bc}$ isoscalar tetraquark turns out to be 
$7.5\times 10^{-15}$\,GeV, in clear disagreement with the result of
 Ref.~\cite{Aga19} obtained using a QCD three-point sum rule approach. 
\begin{table}[b]
\caption{Decay widths, in units of $10^{-15}$\,GeV, for the
processes described by Fig.~\ref{Fig5}.}
  \begin{tabular}{lp{1cm}c}\hline\hline
	Final state & &$\Gamma \, [10^{-15}\,\rm{GeV}]$ \\
	\hline
 $T_{bc} \, e^- \, \nu_e$	 && $3.06\pm0.03$\\
 $T_{bc} \, \mu^- \, \nu_\mu$	     && $3.02\pm0.02$\\
 $T_{bc} \, \tau^- \, \nu_\tau$        && $1.40\pm0.01$\\
				    \hline\hline
  \end{tabular}
	\label{Tab7}
\end{table}

The total decay width of  the $T^-_{bb}$ tetraquark, as calculated in this work, 
is of the order of $\Gamma \approx 87 \times 10^{-15}$\,GeV, which means 
a lifetime $\tau \approx$ 7.6 ps. This lifetime is one order of magnitude larger than the 
simplest guess-by-analogy estimation of $0.3$\,ps of Ref.~\cite{Kar17}.

\section{Summary and outlook}\label{secIV}
We have presented the first comprehensive study of the flavor-exotic $J^P=1^+$ $T_{bb}^-$ 
isoscalar tetraquark. It includes an accurate solution of the four-body problem within a 
quark model, which characterizes the structure of the state,
and an estimate of the lifetime and of the rates 
for the leading semileptonic and nonleptonic decay modes
which are the most promising final states where the tetraquark should be looked for. 
We have shown how pairwise interactions based on color-octet exchange
induce mixing between the $\bar33$ and $6\bar6$ states in the $QQ-\bar q\bar q$ basis,
enhancing the $\bar33$ components for larger values of $M_Q$ due to the
attractive chromoelectric interaction of the $QQ$ pair that it is absent
in the $Q\bar q$ threshold. This result is only valid in the bottom sector. 
In the charm sector, the binding mechanism is different: the $\bar33$
and $6\bar6$ components have a similar probability and are mixed by the 
chromomagnetic interaction. We have  shown how the structure of the $T_{QQ}$ 
state evolves from a molecular-like system to a compact-like structure when 
moving from the charm to the bottom sector.

For the first time, the lifetime of the $T^-_{bb}$ tetraquark has been  calculated
in a quark model beyond simple guess-by-analogy estimations. The total decay width of the 
$T^-_{bb}$ found in this work is  $\Gamma \approx 87 \times 10^{-15}$\,GeV, corresponding to 
a lifetime $\tau \approx 7.6$\,ps. The promising final states are $\bar B^*{}^{-} \, D^*{}^{+} \, l^- \, \bar \nu_\ell$
and $\bar B^*{}^{0} \, D^*{}^{0} \,\ell^- \, \bar \nu_\ell $ among the semileptonic decays, and 
$\bar B^*{}^{-} \, D^*{}^{+} \, D_s^*{}^-$, $\bar B^*{}^{0} \, D^*{}^{0} \, D_s^*{}^- $, and 
$B^*{}^{-} \,D^*{}^{+} \, \rho^-$ among the nonleptonic ones. The 
$T_{bc}^0 \ell^-\nu_\ell$ semileptonic decay is also relevant but in our calculation is not dominant.

Our study complements recent estimates for the production cross sections of $T_{bb}$ 
tetraquarks based on Monte Carlo event generators pointing towards an excellent discovery 
potential in ongoing and forthcoming proton-proton collisions at the LHC~\cite{Ali18}. 
The possible formation of this state in relativistic heavy-ion collisions at the LHC
has also been recently discussed in detail within the quark coalescence model using 
realistic model wave functions with good prospects~\cite{Hon18}.

The spectroscopy of exotic states with hidden heavy flavor has revealed how interesting the 
interaction of heavy hadrons is, with presumably a long-range part of Yukawa type, and a 
short-range part mediated by quark-quark and quark-antiquark forces. A new sector with 
stable flavor-exotic states, such as the $T_{bb}$, remains to be investigated. 
An experimental effort towards the detection of this compact tetraquark states 
is now timely. Its existence is essential to validate our understanding of 
low-energy QCD in the multiquark sector.\\ 

 A long lifetime for the $T_{bb}^-$ tetraquark 
     can ease its detection through the method of ``displaced vertex'' 
     proposed in \cite{Ger18}.\footnote{ We thank A.~Ali for calling
      our attention on this article.}
      
\section{acknowledgments}
This work has been funded by Ministerio de Econom\'\i a, Industria y Competitividad 
and EU FEDER under Contracts No. FPA2016-77177 and FIS2017-84038-C2-1-P, and 
by the EU STRONG-2020 
project under the program H2020-INFRAIA-2018-1, grant agreement no. 824093.

\end{document}